\documentclass[%
 reprint,
superscriptaddress,
 amsmath,amssymb,
aps,
]{revtex4-2}

\usepackage{graphicx}% Include figure files
\usepackage{subcaption}
\usepackage{dcolumn}% Align table columns on decimal point
\usepackage{bm}% bold math
\usepackage{amsmath}
\usepackage{gensymb}
\usepackage{xcolor}
\usepackage{booktabs}
\usepackage{array}
\usepackage{makecell}
\usepackage[figurename=Figure., tablename=Table]{caption}
\newcolumntype{C}[1]{>{\centering\arraybackslash}m{#1}}
\newcommand{\plotsize}{0.32}

\begin{document}

\preprint{APS/123-QED}

\title{A New Workflow for Materials Discovery Bridging the Gap Between Experimental Databases and Graph Neural Networks}% Force line breaks with \\

\author{Brandon Schoener}
\thanks{These authors contributed equally.}
\affiliation{
Department of Physics, University at Buffalo, State University of New York, Buffalo, NY
}

\author{Yuting Hu}
\thanks{These authors contributed equally.}
\affiliation{Department of Computer Science \& Engineering, University at Buffalo, State University of New York, Buffalo, NY}
\affiliation{Institute for Artificial Intelligence and Data Science, University at Buffalo, State University of New York, Buffalo, NY}
\author{Pasit Wanlapha}
\affiliation{Department of Computer Science \& Engineering, University at Buffalo, State University of New York, Buffalo, NY}
\affiliation{Institute for Artificial Intelligence and Data Science, University at Buffalo, State University of New York, Buffalo, NY}
\author{Akshay Rengarajan}
\affiliation{Department of Computer Science \& Engineering, University at Buffalo, State University of New York, Buffalo, NY}
\affiliation{Institute for Artificial Intelligence and Data Science, University at Buffalo, State University of New York, Buffalo, NY}
\author{Ian Moog}
\affiliation{
Department of Physics, University at Buffalo, State University of New York, Buffalo, NY
}
\author{Michael Wang}
\affiliation{
Department of Physics, University at Buffalo, State University of New York, Buffalo, NY
}
\author{Peihong Zhang}
\affiliation{
Department of Physics, University at Buffalo, State University of New York, Buffalo, NY
}
\author{Jinjun Xiong}
\email{jinjun@buffalo.edu}
\affiliation{Department of Computer Science \& Engineering, University at Buffalo, State University of New York, Buffalo, NY}
\affiliation{Institute for Artificial Intelligence and Data Science, University at Buffalo, State University of New York, Buffalo, NY}
\author{Hao Zeng}
\email{haozeng@buffalo.edu}
\affiliation{
Department of Physics, University at Buffalo, State University of New York, Buffalo, NY
}

\begin{abstract}
Incorporating Machine Learning (ML) into material property prediction has become a crucial step in accelerating materials discovery. A key challenge is the severe lack of training data, as many properties are too complicated to calculate with high-throughput first principles techniques. To address this, recent research has created experimental databases from information extracted from scientific literature. However, most existing experimental databases do not provide full atomic coordinate information, which prevents them from supporting advanced ML architectures such as Graph Neural Networks (GNNs). In this work, we propose to bridge this gap through an alignment process between experimental databases and Crystallographic Information Files (CIF) from the Inorganic Crystal Structure Database (ICSD). Our approach enables the creation of a database that can fully leverage state-of-the-art model architectures for material property prediction. It also opens the door to utilizing transfer learning to improve prediction accuracy. To validate our approach, we align NEMAD with the ICSD and compare models trained on the resulting database to those trained on NEMAD originally. We demonstrate significant improvements in both Mean Absolute Error (MAE) and Correct Classification Rate (CCR) in predicting the ordering temperatures and magnetic ground states of magnetic materials, respectively.
\end{abstract}

%\keywords{Suggested keywords}%Use showkeys class option if keyword
                              %display desired
\maketitle
%\tableofcontents

\section{\label{sec:level1}Introduction}
The application of machine learning (ML) techniques to materials discovery has become increasingly important, driven by the need to accelerate property prediction, optimize material design, and uncover hidden structure–property relationships beyond conventional approaches \cite{PhysRevMaterials.2.120301,https://doi.org/10.1002/inf2.12028,https://doi.org/10.1002/pssb.202000600,D0NA00388C,Merchant2023,Choudhary2021,ZHANG2020166998,PhysRevMaterials.8.104404,https://doi.org/10.1002/advs.201900808,Butler2018,Ramprasad2017,Cheng2026}. While first-principles techniques such as Density Functional Theory (DFT) are incredibly powerful for materials prediction, they often carry a trade-off between efficiency and accuracy \cite{PhysRev.140.A1133, Jain2016}. This is where ML has great potential, since the computational cost comes mostly from training. This means using well-trained ML models for materials predictions can achieve DFT level of accuracy while operating at orders of magnitude higher speed and significantly lower cost \cite{PhysRevLett.120.145301,Choudhary2021}. However, in practice this is often limited by the availability of high-quality training data. 

\begin{figure*}[ht!]
    \centering
    \includegraphics[width=1\linewidth]{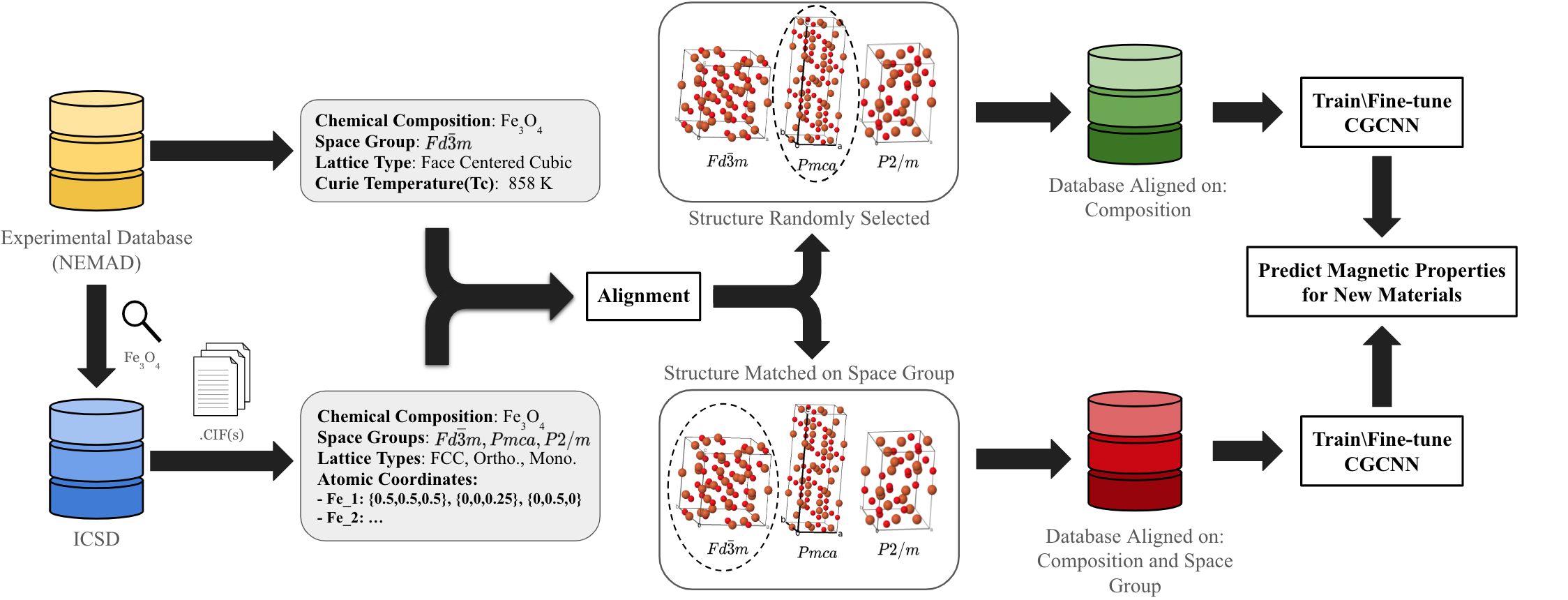}
    \caption{Schematic representing our alignment process followed by subsequent training of CGCNNs. As can be seen above, the database aligned only on composition is prone to align CIFs with incorrect space groups to NEMAD entries.}
    \label{fig:placeholder}
\end{figure*}

Attempts have been made to alleviate the problem by the use of high-throughput first principles calculations to generate computational databases \cite{10.1063/1.4812323, PhysRevMaterials.4.114408, CURTAROLO2012218, Saal2013, Choudhary2020, Haastrup_2018}. Resources such as the Materials Project (MP) or Computational 2D Materials Database use these techniques to efficiently calculate certain properties for a large number of theoretical materials \cite{10.1063/1.4812323,Haastrup_2018}. This has led to databases on the order of 10,000-100,000 entries for properties such as formation energy, band gap, and magnetic ordering. While this quantity is sufficient for training ML models \cite{Merker2022, Lu2022, Acosta2022}, the list of properties calculated and their reliability are limited, in particular for magnetic materials. High-throughput DFT methods do not have the capability to calculate magnetic properties such as magnetocrystalline anisotropy and ordering temperatures to a reliable level of accuracy \cite{PhysRevB.70.132414, PhysRevB.96.134426, Pakdel2025, PhysRevLett.120.097202}. In fact, in their own documentation the MP states that the calculated magnetic ordering should not be taken as a prediction of the true ground state \cite{10.1063/1.4812323}. Instead, such properties are either calculated through models fit to empirical data \cite{10.1063/1.325008}, estimated from DFT‑derived exchange parameters using Mean‑Field Theory \cite{LIECHTENSTEIN198765}, random phase approximation \cite{PhysRevB.64.174402}, or Monte Carlo simulations \cite{PhysRevB.43.6087}. This means that such computational databases are of limited use in training ML models for directly predicting a large range of properties beyond those like formation energy, such as magnetic properties. Instead, smaller databases sourced from experimental results have been successfully utilized to train ML models on magnetic properties \cite{PhysRevMaterials.3.104405,Court2020,Singh2023,Long03042021,10.1063/5.0156377,Nguyen_2019,10.1063/5.0116650,PhysRevApplied.22.024046,XU2024120026}, but this technique is limited by the size and versatility of such databases. Another alternative approach is to utilize ML models to predict formation energy for finding stable phases, and magnetic properties were still calculated with first principles \cite{PhysRevMaterials.8.104404}.

On the other hand, there is a vast amount of experimental data on a range of different materials, largely stored within tens of thousands of research articles, patents, and technical reports. Very recently, approaches have been developed to extract this information using Large Language Models (LLMs) \cite{Itani2025, Miret2025, Court2018, Gilligan2023, ZHANG2024172001}. The Northeast Magnetic Materials Database (NEMAD) was constructed by taking the text from articles and passing it to LLMs, which then output material properties in a structured format \cite{Itani2025}. While this technique shows promise, the database lacks accurate crystal structure information to train advanced ML models. 
In ML-based material prediction, the model accuracy depends crucially on how well the material’s crystal structure is represented. Since atomic configuration directly governs the resulting physical and chemical properties, leading models in this field rely on inputs which encode not only the distinct properties of individual atoms that make up the composition of a material, but also how they are spatially arranged within the crystal lattice. For experimental databases, the level of detail can vary, and most existing databases only give chemical composition and limited structural information such as space groups and lattice types. This restricts the use of more advanced ML models such as Graph Neural Networks (GNN) represented by Crystal Graph Convolutional Neural Network (CGCNN)\cite{PhysRevLett.120.145301}, where the input is a graph constructed from the precise atomic coordinates of the material. As a result, the NEMAD database~\cite{Itani2025} has primarily been used to train relatively simple ML models such as Random Forest (RF), Extreme Gradient Boosting (XGBoost), and ensemble neural networks (ENN) that rely on elemental proportion vectors derived from chemical compositions and one-hot–encoded crystal system categories as input features, for the classification of nonmagnetic, ferromagnetic, and antiferromagnetic materials, as well as for the prediction of Curie and N\'{e}el temperatures. As these models are only trained on general descriptions of structural information, it inherently limits the prediction accuracy due to the sensitivity of magnetic properties to a materials crystal structure \cite{blundell2001magnetism}.

\renewcommand{\arraystretch}{1.15}
\begin{table*}[ht]
\normalsize
\centering
{\setlength{\tabcolsep}{10pt} % increase column spacing
\resizebox{\textwidth}{!}{%
\begin{tabular}{C{4.5cm} C{2.5cm} C{2cm} C{3cm} C{1.5cm} C{2.2cm}}
\toprule
\textbf{Database Alignment} & \textbf{Curie MAE} & \textbf{N\'{e}el MAE} & \textbf{N\'{e}el MAE (TL)} & \textbf{CCR} & \textbf{Noise ($\epsilon$)} \\ 
\midrule
\protect\makecell[c]{No Alignment}
& 56K & 38K & N/A & 0.90 & N/A \\ 
\hline
\protect\makecell[c]{Composition} 
& 44.7K & 28.9K & 25.6K & 0.93 & 7.51~\AA$^2$ \\
\hline
\protect\makecell[c]{Composition + Space Group} 
& \textbf{37.3K} & \textbf{22.6K} & \textbf{\textcolor{red}{22.0K}} & \textbf{0.95} & 0.85~\AA$^2$ \\
\bottomrule
\end{tabular}}}
\caption{Results of models trained under different alignment criteria. TL denotes transfer learning.}
\end{table*}

To address the limitation, in this paper, we propose a new workflow to construct an experimental materials database by aligning existing databases with Crystallographic Information Files (CIF) from the Inorganic Crystal Structure Database (ICSD) \cite{Zagorac:in5024}. A CIF is a file containing the necessary information to completely define a material \cite{hall1991crystallographic}. In particular, it provides the space group representing the symmetry of the system, the unit cell parameters $(a,b,c,\alpha,\beta,\gamma)$ which describe the parallelepiped that is repeated in all directions, and the coordinates of the atoms within the unit cell. The ICSD is a collections of such CIFs which were generated from measurements such as X-Ray or Neutron Diffraction, and are thus representations of physically realizable materials \cite{Zagorac:in5024}. In our aligned database, these CIFs will be mapped to particular magnetic properties based on matches between their complete structural information and the partial structural information from entries in NEMAD. With the aligned database, we train CGCNNs from scratch to predict the ordering temperatures and magnetic orderings of magnetic materials. Our results show significant improvements in Mean Absolute Error (MAE) and Correct Classification Rate (CCR) over those trained on NEMAD using traditional ML models. This underscores the crucial need to utilize advanced model architectures that can handle complete material structure information for accurate material property prediction. Moreover, we further improve the prediction accuracy with transfer learning, where a pre-trained model is fine-tuned with our aligned database \cite{Magar2022, doi:10.1021/acscentsci.9b00804,10.1093/nsr/nwaf066, Jha2019, 10.1063/5.0047066, D3DD00030C}. To verify the quality of our models, we utilize the hand-curated magnetic material database MagNData as a benchmark \cite{Gallego:ks5532}. Our results demonstrate that a highly effective strategy for predicting magnetic properties is to integrate an experimentally curated database with complete structural information obtained through our alignment process, and then to apply advanced ML models such as CGCNN that can take advantage of such structural information.

\section{Methodology}
\subsection{Database Curation}
The workflow for creating our aligned databases is summarized in Figure 1. To curate the databases, we first retrieve crystal structure information from ICSD in CIF format using the chemical compositions reported in our chosen experimental materials database, which in this case is NEMAD. Each ICSD CIF is processed to extract a reduced chemical formula and the space group International Tables (IT) number. We then align ICSD structures with NEMAD magnetic-property entries by matching reduced chemical formulas normalized using PyMatGen. Space group numbers from NEMAD are subsequently compared with those extracted from ICSD and used to assess the consistency of each match, enabling the construction of an aligned subset with improved structural agreement. If one CIF structure matches multiple NEMAD rows, specifically, the same reduced formula but different space groups or magnetic properties then one is randomly chosen to act as the true match. While this process does not guarantee a perfect match, the use of the ICSD improves the chance of a meaningful connection between CIF and NEMAD entry due to the limited number of physically realizable phases for any given composition.
\looseness=-1

Through our alignment process we constructed two databases. By matching entries using chemical composition alone, we obtained Database 1, which contains 11,292 entries with N\'{e}el temperatures and 8,213 entries with Curie temperatures. By matching using both composition and space group, we obtained Database 2, which contains 5,147 entries with N\'{e}el temperatures and 3,821 entries with Curie temperatures. Since CIFs from the ICSD may contain the same chemical composition and even the same space group, multiple CIFs can also be matched with the same NEMAD entry. This means there is structural ambiguity inherent in the alignment process. To evaluate the impact of this ambiguity on the performance of models trained on these databases, we define a noise metric for our databases. This noise is based on differences between the metric tensors of the Niggli reduced forms of CIFs matching the same NEMAD entry \cite{Krivy:a12875}: 

\begin{equation}
    \boldsymbol{G} = 
    \begin{pmatrix}
       a^2 & ab\cos(\gamma) & ac\cos(\beta) \\
       ab\cos(\gamma) & b^2 & bc\cos(\alpha) \\
       ac\cos(\beta) & bc\cos(\alpha) & c^2
    \end{pmatrix}
\end{equation}

Where the Niggli reduced form is a standardized representation of a materials unit cell, and the metric tensor is a a 2x2 matrix which encodes all information about the unit cell of a material. We then define noise as:

\begin{equation}
    \epsilon=\frac{1}{n}\sum_i^n\frac{1}{m_i}\sum_j^{m_i}\lVert{\boldsymbol{G}_{ij}-\hat{\boldsymbol{G}}_i\rVert}
\end{equation}

Where $n$ is the number of NEMAD entries matched to at least one CIF from the ICSD, $m_i$ is the number of CIFs matched to the $i^{th}$ NEMAD entry, $\boldsymbol{G}_{ij}$ is the metric tensor for the $j^{th}$ CIF matched to the $i^{th}$ NEMAD entry, and $\hat{\boldsymbol{G}}_i$ is the average metric tensor matched to the $i^{th}$ NEMAD entry. This quantity captures the average variance of the unit cells matched to any given NEMAD entry in a database.

\subsection{GNNs for Material Property Prediction}
With the complete crystal structures available in the aligned database, we train GNN models to predict Curie temperature, N\'{e}el temperatures, and magnetic ordering. We adopt the CGCNN~\cite{PhysRevLett.120.145301} as the backbone model. Following the graph construction scheme in CGCNN, each crystal structure is represented as a graph in which nodes correspond to atoms in the unit cell and edges are formed between neighboring atoms within a cutoff radius. Each node is associated with a feature vector describing atomic properties, while each edge encodes the interatomic distance between the connected atom pair.

\begin{figure*}[t!]
\captionsetup[sub]{margin={0.35cm, 0cm}}
\begin{subfigure}[t]{\plotsize\textwidth}
  \centering
  \includegraphics[width=0.95\linewidth]{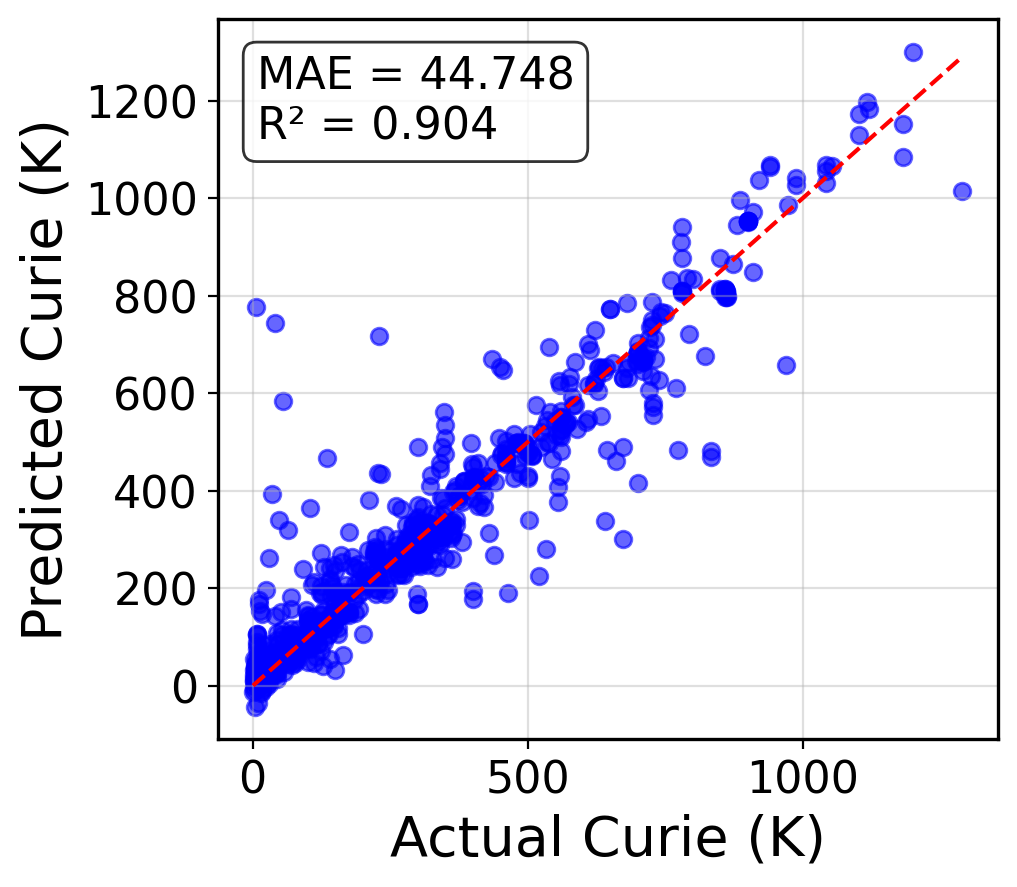}
  \caption{
  Training: \textbf{From Scratch} Alignment:\textbf{Composition}}
  \label{fig:sfig1}
\end{subfigure}
\begin{subfigure}[t]{\plotsize\textwidth}
  \centering
  \includegraphics[width=1\linewidth]{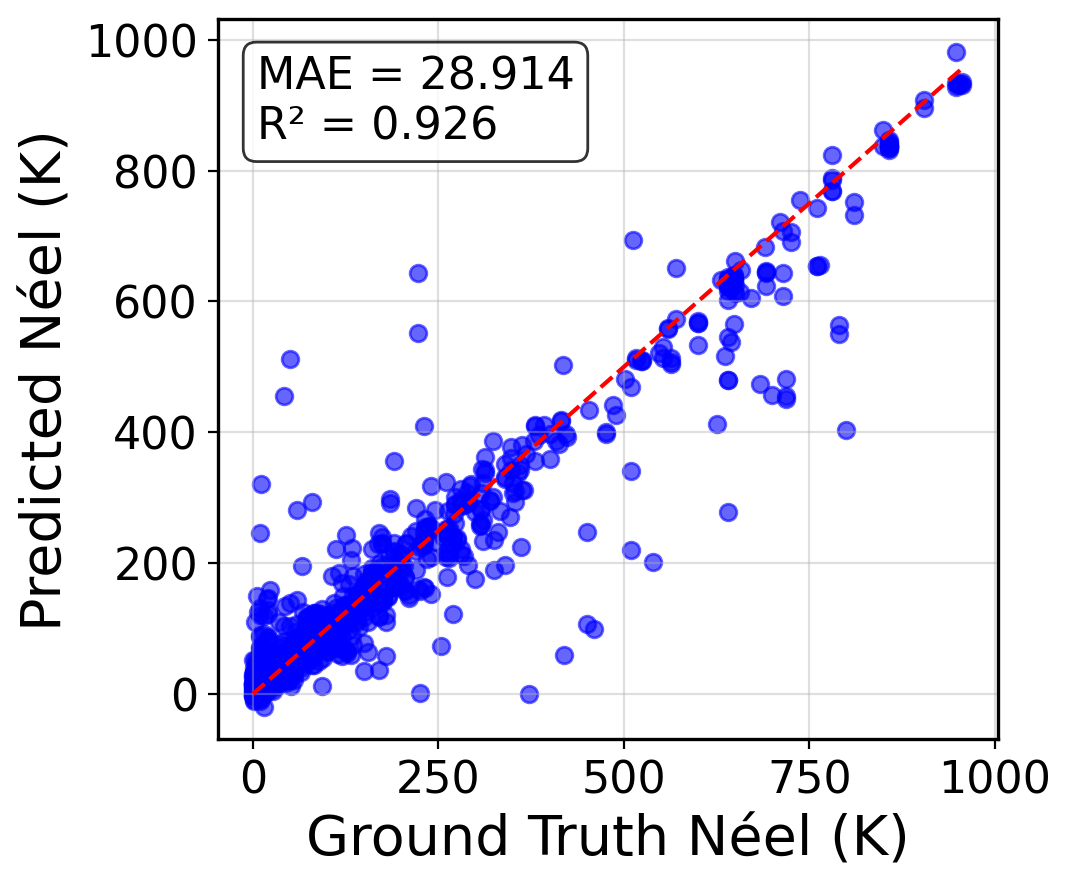}
  \caption{\centering
  Training: \textbf{From Scratch} Alignment: \textbf{Composition}}
  \label{fig:sfig1}
\end{subfigure}
\begin{subfigure}[t]{\plotsize\textwidth}
  \centering
  \includegraphics[width=1\linewidth]{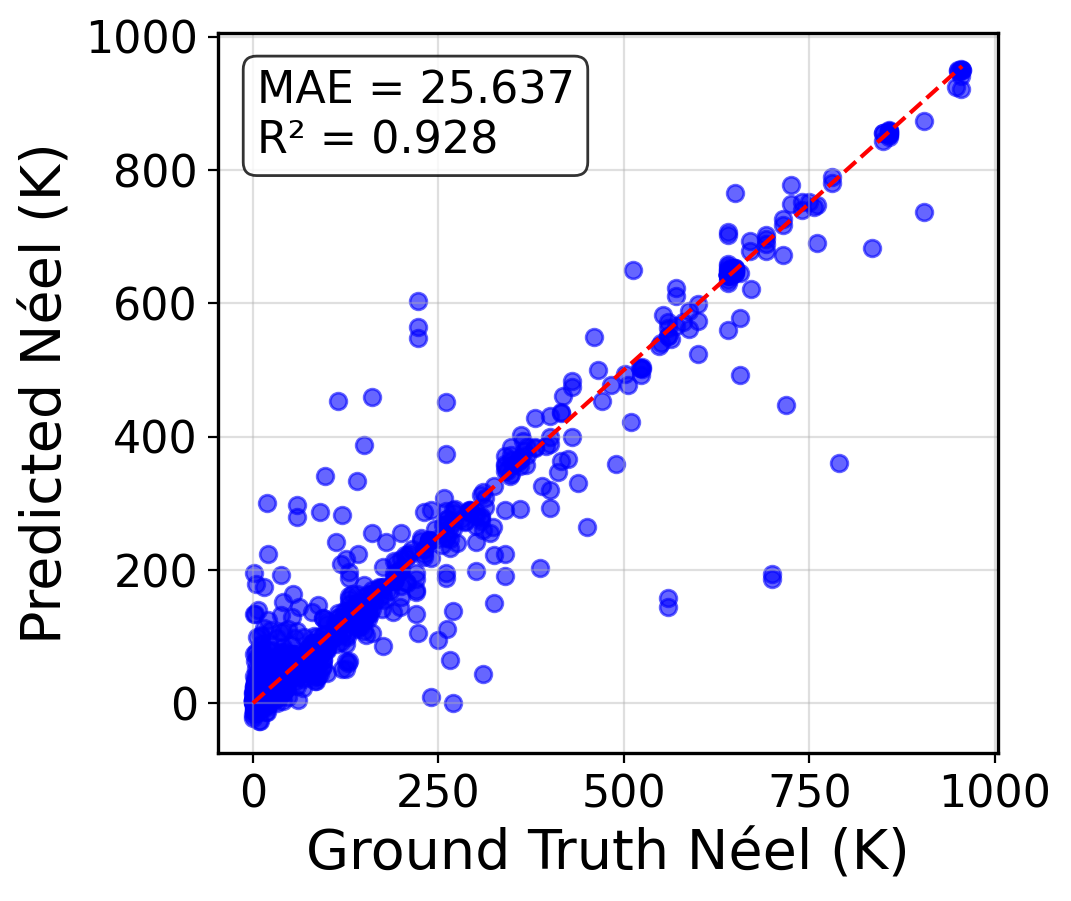}
  \caption{\centering
  Training: \textbf{Fine Tuning} Alignment: \textbf{Composition}}
  \label{fig:sfig1}
\end{subfigure}
\begin{subfigure}[t]{\plotsize\textwidth}
  \centering
  \includegraphics[width=0.95\linewidth]{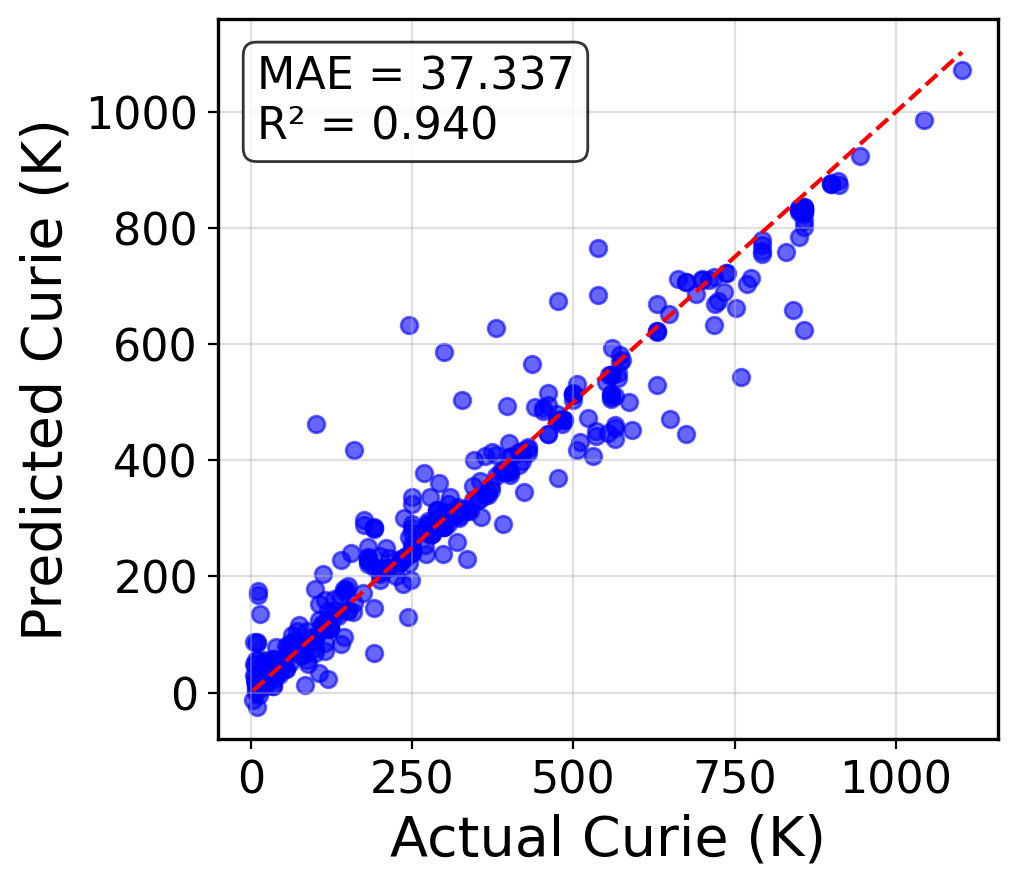}
  \caption{\centering
  Training: \textbf{From Scratch} Alignment: \textbf{Composition + Space Group}}
  \label{fig:sfig2}
\end{subfigure}
\begin{subfigure}[t]{\plotsize\textwidth}
  \centering
  \includegraphics[width=1\linewidth]{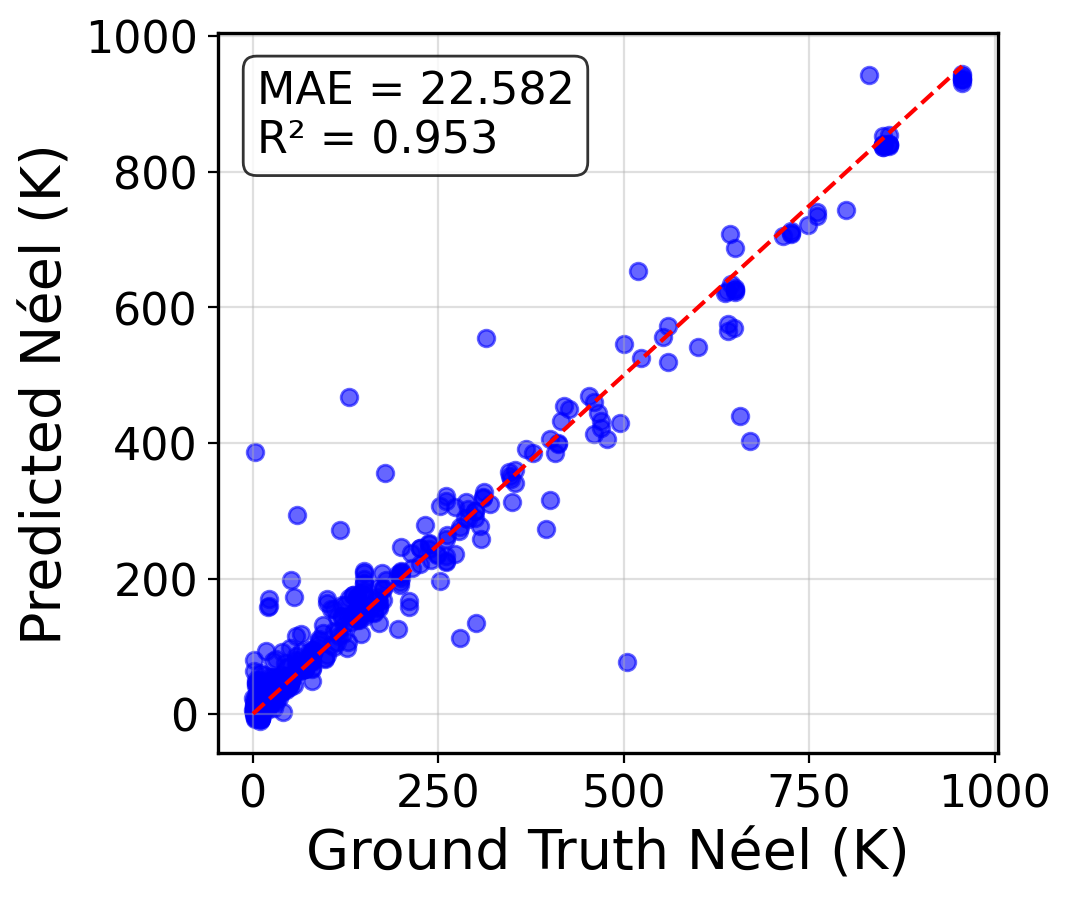}
  \caption{\centering
  Training: \textbf{From Scratch} Alignment: \textbf{Composition + Space Group}}
  \label{fig:sfig1}
\end{subfigure}
\begin{subfigure}[t]{\plotsize\textwidth}
  \centering
  \includegraphics[width=1\linewidth]{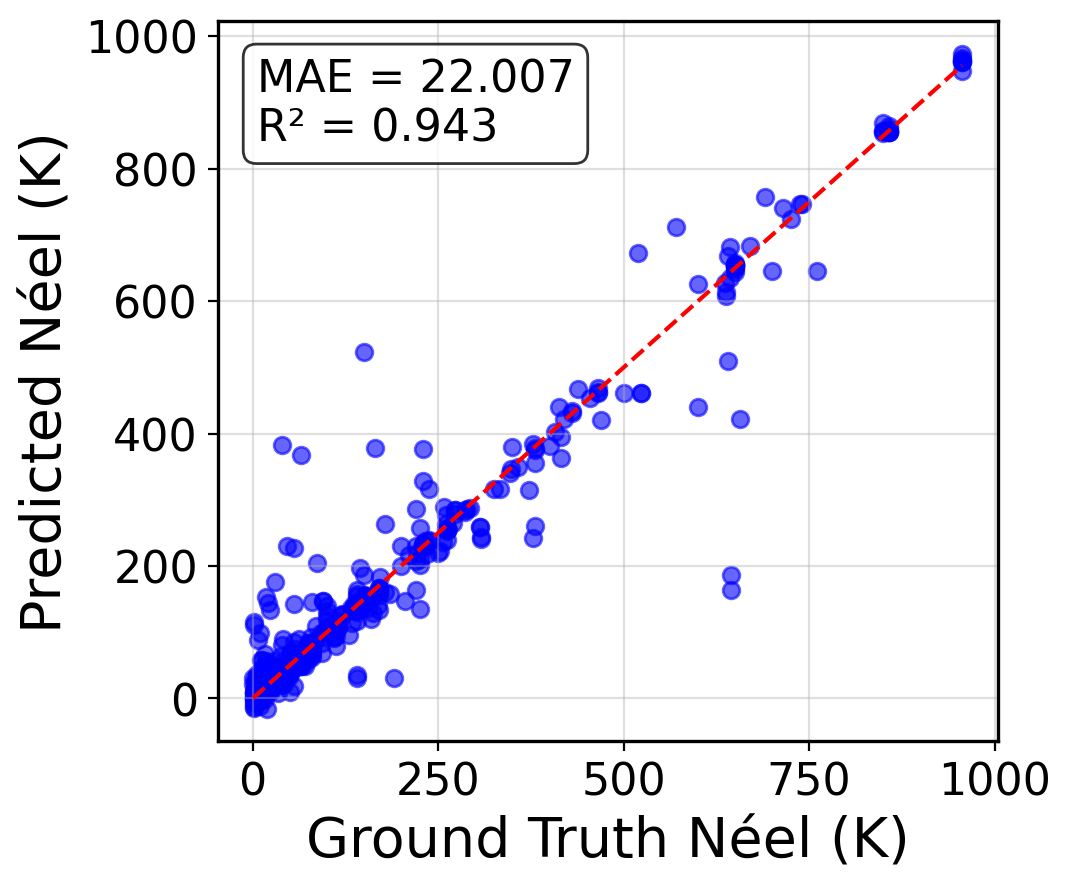}
  \caption{\centering
  Training: \textbf{Fine Tuning} Alignment: \textbf{Composition + Space Group}}
  \label{fig:sfig2}
\end{subfigure}

\caption{Plots of the ground truth N\'{e}el/Curie temperature vs. model prediction for models trained from scratch on databases aligned with composition alone, and aligned with both composition and space group. For N\'{e}el temperature predictions, we also include the results from the models fine-tuned from those trained on formation energy.}
\label{fig:fig}
\end{figure*}

 During the convolution step, these node and edge features are iteratively updated, with each node’s representation transformed based on its own features and those of neighboring nodes, weighted by the edge features connecting them. The updated node feature vectors are then averaged to become a single feature vector through a process called pooling. This new feature vector is then passed through hidden layers in standard Feed Forward Network (FFN) fashion. Extensions have been made to the CGCNN to better extract deep material-property relationships, such as in the iCGCNN where edge features go beyond just distance by incorporating Voronoi‑based attributes, such as facet solid angle, area, and cell volume, enabling richer encoding of the crystal’s local geometry \cite{PhysRevMaterials.4.063801}. The software created for training models from scratch is an extension of that from Ref.~\cite{PhysRevLett.120.145301}. In our implementation, the input graph has node feature vectors of length 64. We set 3 convolutional layers with 128 hidden dimension. A graph pooling operation is applied at the readout for regression tasks. Finally, we pool the output node feature vectors of the final convolutional layer and pass the resulting feature vector through a single output layer, which either gives one (regression) or two (classification) numerical values as our final result. 

To demonstrate the impact of complete structural information on prediction accuracy, we consider N\'{e}el Temperature, Curie Temperature, and Magnetic Ordering prediction tasks under two experimental settings: (1) training CGCNN models from scratch on the aligned database, and (2) transfer learning via fine-tuning a pre-trained CGCNN model using the aligned database. In the second setting, the pre-trained model has already learned to map crystal structures to expressive graph representations that capture essential structural and chemical patterns, providing a strong initialization for predicting the target magnetic properties.

\begin{figure*}[ht]
\centering
\captionsetup[sub]{margin={0.5cm,0cm}}

\begin{subfigure}[t]{0.24\textwidth}
  \centering
  \includegraphics[width=\linewidth]{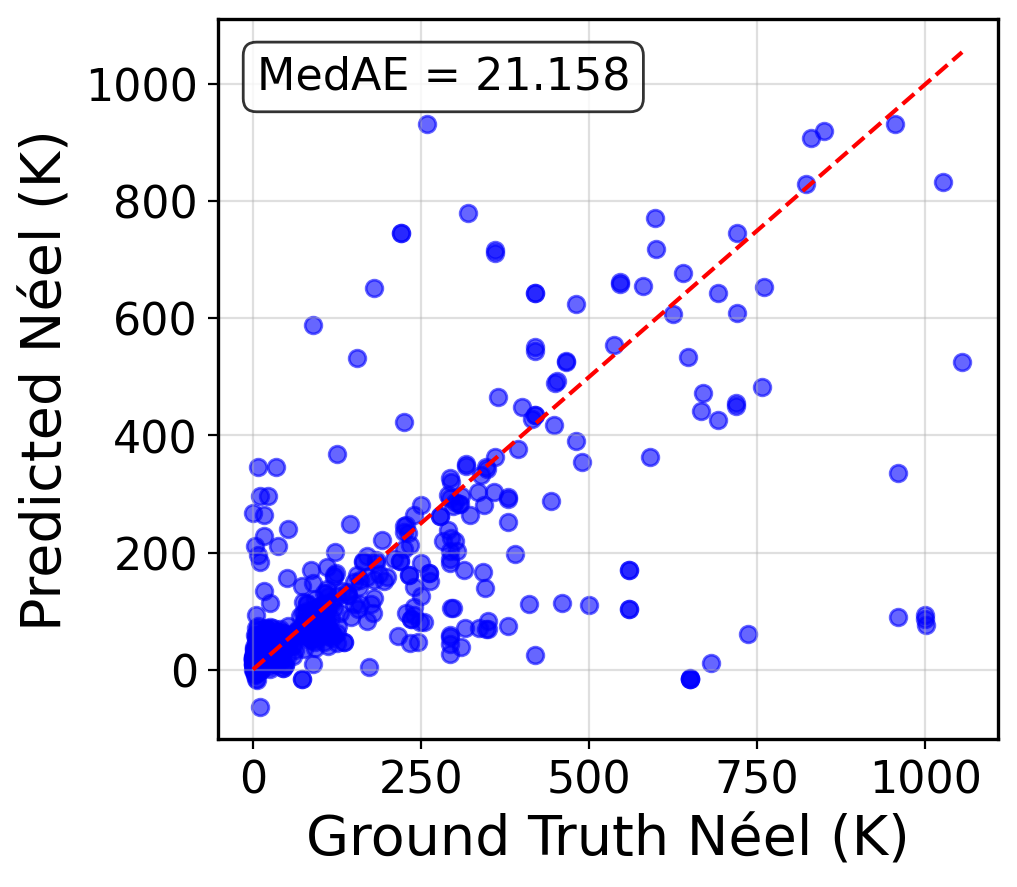}
  \caption{\centering
  Training: \textbf{From Scratch}\\
  Alignment:\\
  \textbf{Composition}}
  \label{fig:sl_comp}
\end{subfigure}\hfill
\begin{subfigure}[t]{0.24\textwidth}
  \centering
  \includegraphics[width=\linewidth]{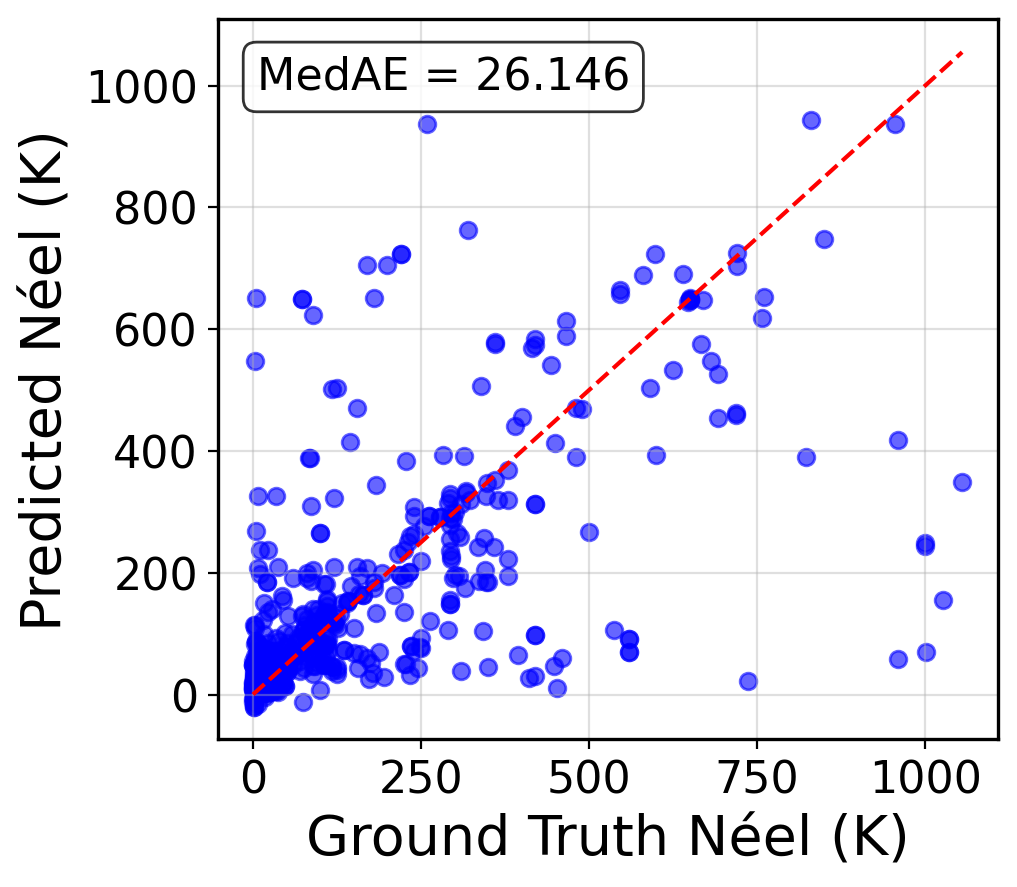}
  \caption{\centering
  Training: \textbf{From Scratch} \\Alignment: \\ \textbf{Composition + Space Group}}
  \label{fig:sl_comp_sg}
\end{subfigure}\hfill
\begin{subfigure}[t]{0.24\textwidth}
  \centering
  \includegraphics[width=\linewidth]{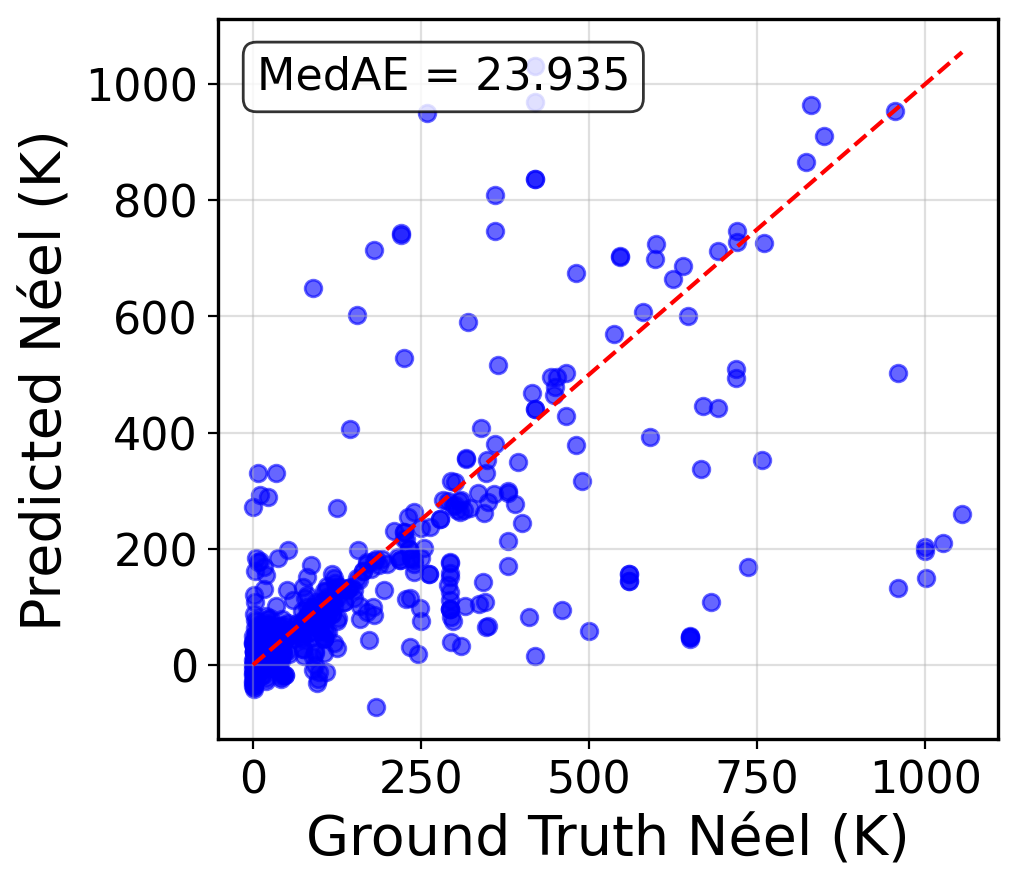}
  \caption{\centering
  Training: \textbf{Fine Tuning} \\Alignment: \\ \textbf{Composition}}
  \label{fig:tl_comp}
\end{subfigure}\hfill
\begin{subfigure}[t]{0.24\textwidth}
  \centering
  \includegraphics[width=\linewidth]{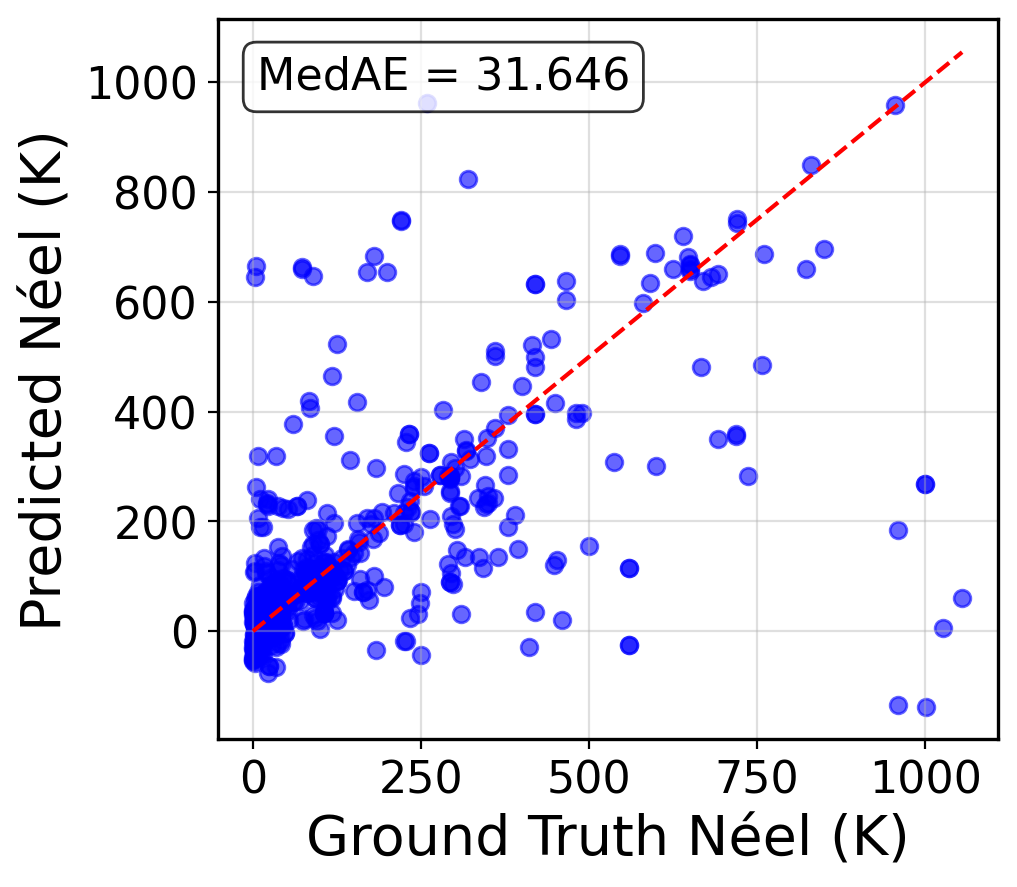}
  \caption{\centering
  Training: \textbf{Fine Tuning} \\Alignment: \\ \textbf{Composition + Space Group}}
  \label{fig:tl_comp_sg}
\end{subfigure}
\caption{Plots of ground truth N\'{e}el temperature vs. model prediction on the MagNData database using models trained from scratch and fine-tuned from those trained on formation energy, along with the MedAE value for each. We show results for both models trained on the database aligned only by composition, and models trained on the database aligned by composition and space group.}
\label{fig:fig}
\end{figure*}

For the model training, the database is divided into training, validation, and test sets using an 80/10/10 split. We optimize the model using the Adam optimizer with an initial learning rate of 0.001, which is decayed to zero via cosine annealing. Training is conducted for up to 300 epochs, with early stopping applied if the validation loss does not improve for 20 consecutive epochs. These choices balance training stability and generalization performance, and they follow commonly adopted best practices in materials informatics models to avoid overfitting while ensuring efficient convergence. In experiment (2), we use a pretrained CGCNN model from Ref.~\cite{PhysRevLett.120.145301}, which is originally designed for formation energy prediction. We only fine-tune its output layer, and follow the same training procedure as experiment (1); we set the starting learning rate for the pre-trained layers to one tenth of that for the output layer.

\section{Experimental Results}
We report model prediction accuracy using databases constructed under different alignment criteria in Table 1. Specifically, we compare the prediction performance of models trained from scratch and fine-tuned using: (i) the original NEMAD database without alignment, (ii) the database aligned by chemical composition, and (iii) the database aligned by both chemical composition and space group. For the evaluation, we employ mean absolute error (MAE) and R$^2$ metrics for regression tasks, and correct classification rate (CCR) metrics for the classification tasks. MAE is calculated by taking the average absolute difference between our predictions and ground truth values, while R$^2$ is calculated as $R^2=1-\frac{\sum_{i=1}^{n} (y_i-\hat{y}_i)}{\sum_{i=1}^{n} (y_i-\bar{y})}$, where $y_i$ and $\hat{y}_i$ are the $i^{th}$ target and predicted values respectively and $\bar{y}$ is the average target value. CCR is calculated as the ratio of correct classifications to the total number of classifications made.

As shown in Table 1, the CGCNNs trained with the aligned database significantly outperform those models without complete material structure input in all metrics \cite{Itani2025}, highlighting the importance of precise crystal structure information for material property predictions. Moreover, we observe further accuracy improvements with transfer learning, which underscores the importance of leveraging pre-trained models with intrinsic built-in representation capability, particularly when the available database is small in scale~\cite{doi:10.1021/acscentsci.9b00804}. Plots of the predicted Curie/N\'{e}el temperatures versus the actual temperatures in every experiment are provided in Figure 2. For each, we see an improvement in both MAE and R$^2$ when moving towards stricter alignment policies, as well as tighter grouping around the target = prediction line.

To verify the quality of the CGCNNs trained on our aligned database, we further test with MagNData, which is a hand-curated experimental database with information on spin configurations and transition temperatures for over 2000 materials, making it a reliable benchmark test set \cite{Gallego:ks5532}. Due to most entries in MagNData being antiferromagnetic, we only tested our N\'{e}el temperature predictive models against this database. The result is shown in Figure 3, our models achieve good alignment along the prediction = target line, as well as low median absolute error (MedAE) comparable to their MAE from their training process. We focus on MedAE rather than MAE and R$^2$ due to the limited amount of data, which leads to aggregate statistics such as MAE and R$^2$ being susceptible to outliers. These results suggest that our models will maintain good performance when applied to the prediction of new materials not present in the original database.

It is observed that the fine-tuned model exhibits comparable but slightly worse performance on the MagNData benchmark compared to those trained from scratch, in terms of MedAE. This unexpected outcome could stem from a combination of two factors. A domain mismatch between the pre-training target property (formation energy) and the new target property (N\'{e}el temperature), and the rich physical complexity underpinning the magnetic properties of materials in MagNData. Consequently, the learned representations may not fully capture spin-dependent interactions critical for predicting N\'{e}el temperatures in this case. To address this limitation, future workflows could pre-train models on magnetic-relevant properties such as magnetic ordering, rather than general thermodynamic quantities.

\section{Conclusion}
In this paper, we trained ML models such as CGCNN using an experimental database for the first time for the prediction of magnetic properties. This was made possible through a novel alignment strategy to match curated experimental databases with CIF files from ICSD to obtain exact crystal structure information. Our results demonstrate that this integration of high-quality experimental data with advanced ML architecture significantly improves prediction accuracy, particularly for complex properties such as Curie and N\'{e}el temperatures. Furthermore, we show that leveraging pre-trained models and transfer learning amplifies these gains, especially when working with limited databases. Our work paves the way for accelerated discovery and design of magnetic materials and beyond. 

\begin{acknowledgments}
This work was supported by The University at Buffalo Institute for Artificial Intelligence and Data Science, and NSF DMR-2242796. HZ acknowledges support from the Moti Lal Rustgi Professorship.
\end{acknowledgments}

\bibliography{library}% Produces the bibliography via BibTeX.

@PREAMBLE{
 "\providecommand{\noopsort}[1]{}" 
 # "\providecommand{\singleletter}[1]{#1}%" 
}

@Article{Itani2025,
author={Itani, Suman
and Zhang, Yibo
and Zang, Jiadong},
title={The northeast materials database for magnetic materials},
journal={Nature Communications},
year={2025},
month={Oct},
day={24},
volume={16},
number={1},
pages={9415},
issn={2041-1723},
doi={10.1038/s41467-025-64458-z},
url={https://doi.org/10.1038/s41467-025-64458-z}
}

@article{PhysRevLett.120.145301,
  title = {Crystal Graph Convolutional Neural Networks for an Accurate and Interpretable Prediction of Material Properties},
  author = {Xie, Tian and Grossman, Jeffrey C.},
  journal = {Phys. Rev. Lett.},
  volume = {120},
  issue = {14},
  pages = {145301},
  numpages = {6},
  year = {2018},
  month = {Apr},
  publisher = {American Physical Society},
  doi = {10.1103/PhysRevLett.120.145301},
  url = {https://link.aps.org/doi/10.1103/PhysRevLett.120.145301}
}

@Article{Magar2022,
author={Magar, Rishikesh
and Wang, Yuyang
and Barati Farimani, Amir},
title={Crystal twins: self-supervised learning for crystalline material property prediction},
journal={npj Computational Materials},
year={2022},
month={Nov},
day={10},
volume={8},
number={1},
pages={231},
issn={2057-3960},
doi={10.1038/s41524-022-00921-5},
url={https://doi.org/10.1038/s41524-022-00921-5}
}

@article{10.1093/nsr/nwaf066,
    author = {Gao, Ze-Feng and Qu, Shuai and Zeng, Bocheng and Liu, Yang and Wen, Ji-Rong and Sun, Hao and Guo, Peng-Jie and Lu, Zhong-Yi},
    title = {AI-accelerated discovery of altermagnetic materials},
    journal = {National Science Review},
    volume = {12},
    number = {4},
    pages = {nwaf066},
    year = {2025},
    month = {02},
    issn = {2095-5138},
    doi = {10.1093/nsr/nwaf066},
    url = {https://doi.org/10.1093/nsr/nwaf066},
    eprint = {https://academic.oup.com/nsr/article-pdf/12/4/nwaf066/62054665/nwaf066.pdf},
}

@article{https://doi.org/10.1002/pssb.202000600,
author = {Katsikas, Georgios and Sarafidis, Charalampos and Kioseoglou, Joseph},
title = {Machine Learning in Magnetic Materials},
journal = {physica status solidi (b)},
volume = {258},
number = {8},
pages = {2000600},
doi = {https://doi.org/10.1002/pssb.202000600},
url = {https://onlinelibrary.wiley.com/doi/abs/10.1002/pssb.202000600},
year = {2021}
}

@article{PhysRevMaterials.2.120301,
  title = {Machine learning in materials design and discovery: Examples from the present and suggestions for the future},
  author = {Gubernatis, J. E. and Lookman, T.},
  journal = {Phys. Rev. Mater.},
  volume = {2},
  issue = {12},
  pages = {120301},
  numpages = {15},
  year = {2018},
  month = {Dec},
  publisher = {American Physical Society},
  doi = {10.1103/PhysRevMaterials.2.120301},
  url = {https://link.aps.org/doi/10.1103/PhysRevMaterials.2.120301}
}

@article{https://doi.org/10.1002/inf2.12028,
author = {Wei, Jing and Chu, Xuan and Sun, Xiang-Yu and Xu, Kun and Deng, Hui-Xiong and Chen, Jigen and Wei, Zhongming and Lei, Ming},
title = {Machine learning in materials science},
journal = {InfoMat},
volume = {1},
number = {3},
pages = {338-358},
doi = {https://doi.org/10.1002/inf2.12028},
url = {https://onlinelibrary.wiley.com/doi/abs/10.1002/inf2.12028},
year = {2019}
}

@article{10.1063/1.4812323,
    author = {Jain, Anubhav and Ong, Shyue Ping and Hautier, Geoffroy and Chen, Wei and Richards, William Davidson and Dacek, Stephen and Cholia, Shreyas and Gunter, Dan and Skinner, David and Ceder, Gerbrand and Persson, Kristin A.},
    title = {Commentary: The Materials Project: A materials genome approach to accelerating materials innovation},
    journal = {APL Materials},
    volume = {1},
    number = {1},
    pages = {011002},
    year = {2013},
    month = {07},
    issn = {2166-532X},
    doi = {10.1063/1.4812323},
    url = {https://doi.org/10.1063/1.4812323},
}

@article{Zagorac:in5024,
author = "Zagorac, D. and M{\"{u}}ller, H. and Ruehl, S. and Zagorac, J. and Rehme, S.",
title = "{Recent developments in the Inorganic Crystal Structure Database: theoretical crystal structure data and related features}",
journal = "Journal of Applied Crystallography",
year = "2019",
volume = "52",
number = "5",
pages = "918--925",
month = "Oct",
doi = {10.1107/S160057671900997X},
url = {https://doi.org/10.1107/S160057671900997X},
}

@article{Gallego:ks5532,
author = "Gallego, Samuel V. and Perez-Mato, J. Manuel and Elcoro, Luis and Tasci, Emre S. and Hanson, Robert M. and Momma, Koichi and Aroyo, Mois I. and Madariaga, Gotzon",
title = "{{\it MAGNDATA}: towards a database of magnetic structures. I.The commensurate case}",
journal = "Journal of Applied Crystallography",
year = "2016",
volume = "49",
number = "5",
pages = "1750--1776",
month = "Oct",
doi = {10.1107/S1600576716012863},
url = {https://doi.org/10.1107/S1600576716012863},
}

@article{doi:10.1021/acscentsci.9b00804,
author = {Yamada, Hironao and Liu, Chang and Wu, Stephen and Koyama, Yukinori and Ju, Shenghong and Shiomi, Junichiro and Morikawa, Junko and Yoshida, Ryo},
title = {Predicting Materials Properties with Little Data Using Shotgun Transfer Learning},
journal = {ACS Central Science},
volume = {5},
number = {10},
pages = {1717-1730},
year = {2019},
doi = {10.1021/acscentsci.9b00804},
note ={PMID: 31660440},
URL = {    
        https://doi.org/10.1021/acscentsci.9b00804
},
eprint = {   
        https://doi.org/10.1021/acscentsci.9b00804
}
}

@Article{D0NA00388C,
author ="Cai, Jiazhen and Chu, Xuan and Xu, Kun and Li, Hongbo and Wei, Jing",
title  ="Machine learning-driven new material discovery",
journal  ="Nanoscale Adv.",
year  ="2020",
volume  ="2",
issue  ="8",
pages  ="3115-3130",
publisher  ="RSC",
doi  ="10.1039/D0NA00388C",
url  ="http://dx.doi.org/10.1039/D0NA00388C"}

@article{PhysRevMaterials.8.104404,
  title = {Machine learning-accelerated discovery of iron cobalt phosphides as rare-earth-free magnets},
  author = {Liao, Timothy and Xia, Weiyi and Sakurai, Masahiro and Zhang, Chao and Sun, Huaijun and Wang, Renhai and Ho, Kai-Ming and Wang, Cai-Zhuang and Chelikowsky, James R.},
  journal = {Phys. Rev. Mater.},
  volume = {8},
  issue = {10},
  pages = {104404},
  numpages = {9},
  year = {2024},
  month = {Oct},
  publisher = {American Physical Society},
  doi = {10.1103/PhysRevMaterials.8.104404},
  url = {https://link.aps.org/doi/10.1103/PhysRevMaterials.8.104404}
}

@Article{Merchant2023,
author={Merchant, Amil
and Batzner, Simon
and Schoenholz, Samuel S.
and Aykol, Muratahan
and Cheon, Gowoon
and Cubuk, Ekin Dogus},
title={Scaling deep learning for materials discovery},
journal={Nature},
year={2023},
month={Dec},
day={01},
volume={624},
number={7990},
pages={80-85},
issn={1476-4687},
doi={10.1038/s41586-023-06735-9},
url={https://doi.org/10.1038/s41586-023-06735-9}
}

@Article{Jha2019,
author={Jha, Dipendra
and Choudhary, Kamal
and Tavazza, Francesca
and Liao, Wei-keng
and Choudhary, Alok
and Campbell, Carelyn
and Agrawal, Ankit},
title={Enhancing materials property prediction by leveraging computational and experimental data using deep transfer learning},
journal={Nature Communications},
year={2019},
month={Nov},
day={22},
volume={10},
number={1},
pages={5316},
issn={2041-1723},
doi={10.1038/s41467-019-13297-w},
url={https://doi.org/10.1038/s41467-019-13297-w}
}

@Article{Choudhary2021,
author={Choudhary, Kamal
and DeCost, Brian},
title={Atomistic Line Graph Neural Network for improved materials property predictions},
journal={npj Computational Materials},
year={2021},
month={Nov},
day={15},
volume={7},
number={1},
pages={185},
issn={2057-3960},
doi={10.1038/s41524-021-00650-1},
url={https://doi.org/10.1038/s41524-021-00650-1}
}

@article{Krivy:a12875,
author = "K{\v{r}}iv{\'{y}}, I. and Gruber, B.",
title = "{A unified algorithm for determining the reduced (Niggli) cell}",
journal = "Acta Crystallographica Section A",
year = "1976",
volume = "32",
number = "2",
pages = "297--298",
month = "Mar",
doi = {10.1107/S0567739476000636},
url = {https://doi.org/10.1107/S0567739476000636},
}

@article{PhysRevMaterials.4.063801,
  title = {Developing an improved crystal graph convolutional neural network framework for accelerated materials discovery},
  author = {Park, Cheol Woo and Wolverton, Chris},
  journal = {Phys. Rev. Mater.},
  volume = {4},
  issue = {6},
  pages = {063801},
  numpages = {11},
  year = {2020},
  month = {Jun},
  publisher = {American Physical Society},
  doi = {10.1103/PhysRevMaterials.4.063801},
  url = {https://link.aps.org/doi/10.1103/PhysRevMaterials.4.063801}
}

@article{10.1063/5.0047066,
    author = {Kong, Shufeng and Guevarra, Dan and Gomes, Carla P. and Gregoire, John M.},
    title = {Materials representation and transfer learning for multi-property prediction},
    journal = {Applied Physics Reviews},
    volume = {8},
    number = {2},
    pages = {021409},
    year = {2021},
    month = {06},
    issn = {1931-9401},
    doi = {10.1063/5.0047066}
}

@Article{D3DD00030C,
author ="Hoffmann, Noah and Schmidt, Jonathan and Botti, Silvana and Marques, Miguel A. L.",
title  ="Transfer learning on large datasets for the accurate prediction of material properties",
journal  ="Digital Discovery",
year  ="2023",
volume  ="2",
issue  ="5",
pages  ="1368-1379",
publisher  ="RSC",
doi  ="10.1039/D3DD00030C",
url  ="http://dx.doi.org/10.1039/D3DD00030C"}

@article{PhysRevMaterials.4.114408,
  title = {Discovering rare-earth-free magnetic materials through the development of a database},
  author = {Sakurai, Masahiro and Wang, Renhai and Liao, Timothy and Zhang, Chao and Sun, Huaijun and Sun, Yang and Wang, Haidi and Zhao, Xin and Wang, Songyou and Balasubramanian, Balamurugan and Xu, Xiaoshan and Sellmyer, David J. and Antropov, Vladimir and Zhang, Jianhua and Wang, Cai-Zhuang and Ho, Kai-Ming and Chelikowsky, James R.},
  journal = {Phys. Rev. Mater.},
  volume = {4},
  issue = {11},
  pages = {114408},
  numpages = {15},
  year = {2020},
  month = {Nov},
  publisher = {American Physical Society},
  doi = {10.1103/PhysRevMaterials.4.114408},
  url = {https://link.aps.org/doi/10.1103/PhysRevMaterials.4.114408}
}

@article{ZHANG2020166998,
title = {Curie temperature modeling of magnetocaloric lanthanum manganites using Gaussian process regression},
journal = {Journal of Magnetism and Magnetic Materials},
volume = {512},
pages = {166998},
year = {2020},
issn = {0304-8853},
doi = {https://doi.org/10.1016/j.jmmm.2020.166998},
url = {https://www.sciencedirect.com/science/article/pii/S0304885319339642},
author = {Yun Zhang and Xiaojie Xu}
}

@Article{Miret2025,
author={Miret, Santiago
and Krishnan, N. M. Anoop},
title={Enabling large language models for real-world materials discovery},
journal={Nature Machine Intelligence},
year={2025},
month={Jul},
day={01},
volume={7},
number={7},
pages={991-998},
issn={2522-5839},
doi={10.1038/s42256-025-01058-y},
url={https://doi.org/10.1038/s42256-025-01058-y}
}

@article{CURTAROLO2012218,
title = {AFLOW: An automatic framework for high-throughput materials discovery},
journal = {Computational Materials Science},
volume = {58},
pages = {218-226},
year = {2012},
issn = {0927-0256},
doi = {https://doi.org/10.1016/j.commatsci.2012.02.005},
url = {https://www.sciencedirect.com/science/article/pii/S0927025612000717},
author = {Stefano Curtarolo and Wahyu Setyawan and Gus L.W. Hart and Michal Jahnatek and Roman V. Chepulskii and Richard H. Taylor and Shidong Wang and Junkai Xue and Kesong Yang and Ohad Levy and Michael J. Mehl and Harold T. Stokes and Denis O. Demchenko and Dane Morgan}
}

@Article{Saal2013,
author={Saal, James E.
and Kirklin, Scott
and Aykol, Muratahan
and Meredig, Bryce
and Wolverton, C.},
title={Materials Design and Discovery with High-Throughput Density Functional Theory: The Open Quantum Materials Database (OQMD)},
journal={JOM},
year={2013},
month={Nov},
day={01},
volume={65},
number={11},
pages={1501-1509},
issn={1543-1851},
doi={10.1007/s11837-013-0755-4},
url={https://doi.org/10.1007/s11837-013-0755-4}
}

@Article{Choudhary2020,
author={Choudhary, Kamal
and Garrity, Kevin F.
and Reid, Andrew C. E.
and DeCost, Brian
and Biacchi, Adam J.
and Hight Walker, Angela R.
and Trautt, Zachary
and Hattrick-Simpers, Jason
and Kusne, A. Gilad
and Centrone, Andrea
and Davydov, Albert
and Jiang, Jie
and Pachter, Ruth
and Cheon, Gowoon
and Reed, Evan
and Agrawal, Ankit
and Qian, Xiaofeng
and Sharma, Vinit
and Zhuang, Houlong
and Kalinin, Sergei V.
and Sumpter, Bobby G.
and Pilania, Ghanshyam
and Acar, Pinar
and Mandal, Subhasish
and Haule, Kristjan
and Vanderbilt, David
and Rabe, Karin
and Tavazza, Francesca},
title={The joint automated repository for various integrated simulations (JARVIS) for data-driven materials design},
journal={npj Computational Materials},
year={2020},
month={Nov},
day={12},
volume={6},
number={1},
pages={173},
issn={2057-3960},
doi={10.1038/s41524-020-00440-1},
url={https://doi.org/10.1038/s41524-020-00440-1}
}

@article{https://doi.org/10.1002/advs.201900808,
author = {Himanen, Lauri and Geurts, Amber and Foster, Adam Stuart and Rinke, Patrick},
title = {Data-Driven Materials Science: Status, Challenges, and Perspectives},
journal = {Advanced Science},
volume = {6},
number = {21},
pages = {1900808},
doi = {https://doi.org/10.1002/advs.201900808},
year = {2019}
}

@article{PhysRev.140.A1133,
  title = {Self-Consistent Equations Including Exchange and Correlation Effects},
  author = {Kohn, W. and Sham, L. J.},
  journal = {Phys. Rev.},
  volume = {140},
  issue = {4A},
  pages = {A1133--A1138},
  numpages = {0},
  year = {1965},
  month = {Nov},
  publisher = {American Physical Society},
  doi = {10.1103/PhysRev.140.A1133},
  url = {https://link.aps.org/doi/10.1103/PhysRev.140.A1133}
}

@Article{Jain2016,
author={Jain, Anubhav
and Shin, Yongwoo
and Persson, Kristin A.},
title={Computational predictions of energy materials using density functional theory},
journal={Nature Reviews Materials},
year={2016},
month={Jan},
day={11},
volume={1},
number={1},
pages={15004},
issn={2058-8437},
doi={10.1038/natrevmats.2015.4},
url={https://doi.org/10.1038/natrevmats.2015.4}
}

@article{10.1063/1.325008,
    author = {Mimura, Y. and Imamura, N. and Kobayashi, T. and Okada, A. and Kushiro, Y.},
    title = {Magnetic properties of amorphous alloy films of Fe with Gd, Tb, Dy, Ho, or Er},
    journal = {Journal of Applied Physics},
    volume = {49},
    number = {3},
    pages = {1208-1215},
    year = {1978},
    month = {03},
    issn = {0021-8979},
    doi = {10.1063/1.325008}
}

@Article{Butler2018,
author={Butler, Keith T.
and Davies, Daniel W.
and Cartwright, Hugh
and Isayev, Olexandr
and Walsh, Aron},
title={Machine learning for molecular and materials science},
journal={Nature},
year={2018},
month={Jul},
day={01},
volume={559},
number={7715},
pages={547-555},
issn={1476-4687},
doi={10.1038/s41586-018-0337-2},
}

@Article{Ramprasad2017,
author={Ramprasad, Rampi
and Batra, Rohit
and Pilania, Ghanshyam
and Mannodi-Kanakkithodi, Arun
and Kim, Chiho},
title={Machine learning in materials informatics: recent applications and prospects},
journal={npj Computational Materials},
year={2017},
month={Dec},
day={13},
volume={3},
number={1},
pages={54},
issn={2057-3960},
doi={10.1038/s41524-017-0056-5},
url={https://doi.org/10.1038/s41524-017-0056-5}
}

@article{PhysRevMaterials.3.104405,
  title = {Predicting the Curie temperature of ferromagnets using machine learning},
  author = {Nelson, James and Sanvito, Stefano},
  journal = {Phys. Rev. Mater.},
  volume = {3},
  issue = {10},
  pages = {104405},
  numpages = {11},
  year = {2019},
  month = {Oct},
  publisher = {American Physical Society},
  doi = {10.1103/PhysRevMaterials.3.104405},
  url = {https://link.aps.org/doi/10.1103/PhysRevMaterials.3.104405}
}

@Article{Court2020,
author={Court, Callum J.
and Cole, Jacqueline M.},
title={Magnetic and superconducting phase diagrams and transition temperatures predicted using text mining and machine learning},
journal={npj Computational Materials},
year={2020},
month={Mar},
day={13},
volume={6},
number={1},
pages={18},
issn={2057-3960},
doi={10.1038/s41524-020-0287-8},
url={https://doi.org/10.1038/s41524-020-0287-8}
}

@Article{Singh2023,
author={Singh, Prashant
and Del Rose, Tyler
and Palasyuk, Andriy
and Mudryk, Yaroslav},
title={Physics-Informed Machine-Learning Prediction of Curie Temperatures and Its Promise for Guiding the Discovery of Functional Magnetic Materials},
journal={Chemistry of Materials},
year={2023},
month={Aug},
day={22},
publisher={American Chemical Society},
volume={35},
number={16},
pages={6304-6312},
issn={0897-4756},
doi={10.1021/acs.chemmater.3c00892},
url={https://doi.org/10.1021/acs.chemmater.3c00892}
}

@article{Long03042021,
author = {Teng Long and Nuno M. Fortunato and Yixuan Zhang and Oliver Gutfleisch and Hongbin Zhang},
title = {An accelerating approach of designing ferromagnetic materials via machine learning modeling of magnetic ground state and Curie temperature},
journal = {Materials Research Letters},
volume = {9},
number = {4},
pages = {169--174},
year = {2021},
publisher = {Taylor \& Francis},
doi = {10.1080/21663831.2020.1863876},
URL = {https://doi.org/10.1080/21663831.2020.1863876},
eprint = {https://doi.org/10.1080/21663831.2020.1863876}
}

@article{10.1063/5.0156377,
    author = {Belot, Joshua F. and Taufour, Valentin and Sanvito, Stefano and Hart, Gus L. W.},
    title = {Machine learning predictions of high-Curie-temperature materials},
    journal = {Applied Physics Letters},
    volume = {123},
    number = {4},
    pages = {042405},
    year = {2023},
    month = {07},
    issn = {0003-6951},
    doi = {10.1063/5.0156377}
}

@article{Nguyen_2019,
doi = {10.1088/1742-6596/1290/1/012009},
url = {https://doi.org/10.1088/1742-6596/1290/1/012009},
year = {2019},
month = {oct},
publisher = {IOP Publishing},
volume = {1290},
number = {1},
pages = {012009},
author = {Nguyen, Duong-Nguyen and Pham, Tien-Lam and Nguyen, Viet-Cuong and Nguyen, Anh-Tuan and Kino, Hiori and Miyake, Takashi and Dam, Hieu-Chi},
title = {A regression-based model evaluation of the Curie temperature of transition-metal rare-earth compounds},
journal = {Journal of Physics: Conference Series}
}

@article{10.1063/5.0116650,
    author = {Choudhary, Amit Kumar and Kini, Anoop and Hohs, Dominic and Jansche, Andreas and Bernthaler, Timo and Csiszár, Orsolya and Goll, Dagmar and Schneider, Gerhard},
    title = {Machine learning-based Curie temperature prediction for magnetic 14:2:1 phases},
    journal = {AIP Advances},
    volume = {13},
    number = {3},
    pages = {035112},
    year = {2023},
    month = {03},
    issn = {2158-3226},
    doi = {10.1063/5.0116650}
}

@article{PhysRevApplied.22.024046,
  title = {Accurate machine-learning predictions of coercivity in high-performance permanent magnets},
  author = {Bhandari, Churna and Nop, Gavin N. and Smith, Jonathan D.H. and Paudyal, Durga},
  journal = {Phys. Rev. Appl.},
  volume = {22},
  issue = {2},
  pages = {024046},
  numpages = {22},
  year = {2024},
  month = {Aug},
  publisher = {American Physical Society},
  doi = {10.1103/PhysRevApplied.22.024046},
  url = {https://link.aps.org/doi/10.1103/PhysRevApplied.22.024046}
}

@article{XU2024120026,
title = {Predicting the Curie temperature of Sm-Co-based alloys via data-driven strategy},
journal = {Acta Materialia},
volume = {274},
pages = {120026},
year = {2024},
issn = {1359-6454},
doi = {https://doi.org/10.1016/j.actamat.2024.120026},
url = {https://www.sciencedirect.com/science/article/pii/S1359645424003781},
author = {Guojing Xu and Feng Cheng and Hao Lu and Chao Hou and Xiaoyan Song}
}

@Article{Court2018,
author={Court, Callum J.
and Cole, Jacqueline M.},
title={Auto-generated materials database of Curie and N{\'e}el temperatures via semi-supervised relationship extraction},
journal={Scientific Data},
year={2018},
month={Jun},
day={19},
volume={5},
number={1},
pages={180111},
issn={2052-4463},
doi={10.1038/sdata.2018.111},
url={https://doi.org/10.1038/sdata.2018.111}
}

@Article{Gilligan2023,
author={Gilligan, Luke P. J.
and Cobelli, Matteo
and Taufour, Valentin
and Sanvito, Stefano},
title={A rule-free workflow for the automated generation of databases from scientific literature},
journal={npj Computational Materials},
year={2023},
month={Dec},
day={13},
volume={9},
number={1},
pages={222},
issn={2057-3960},
doi={10.1038/s41524-023-01171-9},
url={https://doi.org/10.1038/s41524-023-01171-9}
}

@book{blundell2001magnetism,
  title={Magnetism in condensed matter},
  author={Blundell, Stephen},
  year={2001},
  publisher={OUP Oxford}
}

@article{ZHANG2024172001,
title = {GPTArticleExtractor: An automated workflow for magnetic material database construction},
journal = {Journal of Magnetism and Magnetic Materials},
volume = {597},
pages = {172001},
year = {2024},
issn = {0304-8853},
doi = {https://doi.org/10.1016/j.jmmm.2024.172001},
url = {https://www.sciencedirect.com/science/article/pii/S0304885324002920},
author = {Yibo Zhang and Suman Itani and Kamal Khanal and Emmanuel Okyere and Gavin Smith and Koichiro Takahashi and Jiadong Zang}
}

@article{PhysRevB.70.132414,
  title = {Extent and limitations of density-functional theory in describing magnetic systems},
  author = {Illas, F. and de P. R. Moreira, I. and Bofill, J. M. and Filatov, M.},
  journal = {Phys. Rev. B},
  volume = {70},
  issue = {13},
  pages = {132414},
  numpages = {4},
  year = {2004},
  month = {Oct},
  publisher = {American Physical Society},
  doi = {10.1103/PhysRevB.70.132414},
  url = {https://link.aps.org/doi/10.1103/PhysRevB.70.132414}
}

@article{Haastrup_2018,
doi = {10.1088/2053-1583/aacfc1},
url = {https://doi.org/10.1088/2053-1583/aacfc1},
year = {2018},
month = {sep},
publisher = {IOP Publishing},
volume = {5},
number = {4},
pages = {042002},
author = {Haastrup, Sten and Strange, Mikkel and Pandey, Mohnish and Deilmann, Thorsten and Schmidt, Per S and Hinsche, Nicki F and Gjerding, Morten N and Torelli, Daniele and Larsen, Peter M and Riis-Jensen, Anders C and Gath, Jakob and Jacobsen, Karsten W and Jørgen Mortensen, Jens and Olsen, Thomas and Thygesen, Kristian S},
title = {The Computational 2D Materials Database: high-throughput modeling and discovery of atomically thin crystals},
journal = {2D Materials}
}

@article{PhysRevB.96.134426,
  title = {Quantifying confidence in density functional theory predictions of magnetic ground states},
  author = {Houchins, Gregory and Viswanathan, Venkatasubramanian},
  journal = {Phys. Rev. B},
  volume = {96},
  issue = {13},
  pages = {134426},
  numpages = {8},
  year = {2017},
  month = {Oct},
  publisher = {American Physical Society},
  doi = {10.1103/PhysRevB.96.134426},
  url = {https://link.aps.org/doi/10.1103/PhysRevB.96.134426}
}

@Article{Pakdel2025,
author={Pakdel, Sahar
and Olsen, Thomas
and Thygesen, Kristian S.},
title={Effect of Hubbard U-corrections on the electronic and magnetic properties of 2D materials: a high-throughput study},
journal={npj Computational Materials},
year={2025},
month={Jan},
day={24},
volume={11},
number={1},
pages={18},
issn={2057-3960},
doi={10.1038/s41524-024-01503-3},
url={https://doi.org/10.1038/s41524-024-01503-3}
}

@article{PhysRevLett.120.097202,
  title = {Calculating the Magnetic Anisotropy of Rare-Earth--Transition-Metal Ferrimagnets},
  author = {Patrick, Christopher E. and Kumar, Santosh and Balakrishnan, Geetha and Edwards, Rachel S. and Lees, Martin R. and Petit, Leon and Staunton, Julie B.},
  journal = {Phys. Rev. Lett.},
  volume = {120},
  issue = {9},
  pages = {097202},
  numpages = {6},
  year = {2018},
  month = {Feb},
  publisher = {American Physical Society},
  doi = {10.1103/PhysRevLett.120.097202},
  url = {https://link.aps.org/doi/10.1103/PhysRevLett.120.097202}
}

@Article{Merker2022,
author={Merker, Helena A.
and Heiberger, Harry
and Nguyen, Linh
and Liu, Tongtong
and Chen, Zhantao
and Andrejevic, Nina
and Drucker, Nathan C.
and Okabe, Ryotaro
and Kim, Song Eun
and Wang, Yao
and Smidt, Tess
and Li, Mingda},
title={Machine learning magnetism classifiers from atomic coordinates},
journal={iScience},
year={2022},
month={Oct},
day={21},
publisher={Elsevier},
volume={25},
number={10},
issn={2589-0042},
doi={10.1016/j.isci.2022.105192},
url={https://doi.org/10.1016/j.isci.2022.105192}
}

@Article{Lu2022,
author={Lu, Shuaihua
and Zhou, Qionghua
and Guo, Yilv
and Wang, Jinlan},
title={On-the-fly interpretable machine learning for rapid discovery of two-dimensional ferromagnets with high Curie temperature},
journal={Chem},
year={2022},
month={Mar},
day={10},
publisher={Elsevier},
volume={8},
number={3},
pages={769-783},
issn={2451-9294},
doi={10.1016/j.chempr.2021.11.009},
url={https://doi.org/10.1016/j.chempr.2021.11.009}
}

@Article{Acosta2022,
author={Acosta, Carlos Mera
and Ogoshi, Elton
and Souza, Jose Antonio
and Dalpian, Gustavo M.},
title={Machine Learning Study of the Magnetic Ordering in 2D Materials},
journal={ACS Applied Materials {\&} Interfaces},
year={2022},
month={Feb},
day={23},
publisher={American Chemical Society},
volume={14},
number={7},
pages={9418-9432},
issn={1944-8244},
doi={10.1021/acsami.1c21558},
url={https://doi.org/10.1021/acsami.1c21558}
}

@article{LIECHTENSTEIN198765,
title = {Local spin density functional approach to the theory of exchange interactions in ferromagnetic metals and alloys},
journal = {Journal of Magnetism and Magnetic Materials},
volume = {67},
number = {1},
pages = {65-74},
year = {1987},
issn = {0304-8853},
doi = {https://doi.org/10.1016/0304-8853(87)90721-9},
url = {https://www.sciencedirect.com/science/article/pii/0304885387907219},
author = {A.I. Liechtenstein and M.I. Katsnelson and V.P. Antropov and V.A. Gubanov}
}

@article{PhysRevB.64.174402,
  title = {Ab initio calculations of exchange interactions, spin-wave stiffness constants, and Curie temperatures of Fe, Co, and Ni},
  author = {Pajda, M. and Kudrnovsk\'y, J. and Turek, I. and Drchal, V. and Bruno, P.},
  journal = {Phys. Rev. B},
  volume = {64},
  issue = {17},
  pages = {174402},
  numpages = {9},
  year = {2001},
  month = {Oct},
  publisher = {American Physical Society},
  doi = {10.1103/PhysRevB.64.174402},
  url = {https://link.aps.org/doi/10.1103/PhysRevB.64.174402}
}

@article{PhysRevB.43.6087,
  title = {High-accuracy Monte Carlo study of the three-dimensional classical Heisenberg ferromagnet},
  author = {Peczak, P. and Ferrenberg, Alan M. and Landau, D. P.},
  journal = {Phys. Rev. B},
  volume = {43},
  issue = {7},
  pages = {6087--6093},
  numpages = {0},
  year = {1991},
  month = {Mar},
  publisher = {American Physical Society},
  doi = {10.1103/PhysRevB.43.6087},
  url = {https://link.aps.org/doi/10.1103/PhysRevB.43.6087}
}

@Article{Cheng2026,
author={Cheng, Mouyang
and Fu, Chu-Liang
and Okabe, Ryotaro
and Chotrattanapituk, Abhijatmedhi
and Boonkird, Artittaya
and Hung, Nguyen Tuan
and Li, Mingda},
title={Artificial intelligence-driven approaches for materials design and discovery},
journal={Nature Materials},
year={2026},
month={Jan},
day={02},
issn={1476-4660},
doi={10.1038/s41563-025-02403-7},
}

@article{hall1991crystallographic,
  title={The crystallographic information file (CIF): a new standard archive file for crystallography},
  author={Hall, Sydney R and Allen, Frank H and Brown, I David},
  journal={Foundations of Crystallography},
  volume={47},
  number={6},
  pages={655--685},
  year={1991},
  publisher={International Union of Crystallography}
}

\end{document}